**Preparing students to be leaders of the quantum information revolution**

Chandralekha Singh[1,2], Abraham Asfaw[3], and Jeremy Levy[1,2]

[1]Department of Physics and Astronomy, University of Pittsburgh, Pittsburgh, PA
[2]Pittsburgh Quantum Institute, Pittsburgh, PA
[3]IBM Quantum, IBM T.J. Watson Research Center, Yorktown Heights, New York, NY

*To meet future research and workforce demands, the physics community needs to embrace the challenge of educating students with diverse educational backgrounds.*

As the crowning technological inventions of the first quantum revolution—transistors, lasers, and computers—continue to enrich our lives, newfound excitement surrounds the use of quantum phenomena to create a second quantum revolution. Quantum computers will compute faster than existing classical computers and enable computations that were not previously possible. Quantum sensors will detect one part in a million variations in Earth's gravitational field or tiny magnetic fields emanating from the human brain. Quantum communication technologies will send information securely over large distances, protected by fundamental laws of nature.

Those technologies could dramatically improve society. But the current educational system isn't prepared to meet the surging demand for workers, researchers, and teachers that understand and can teach the core concepts in quantum information science and technology (QIST), the field from which they have emerged. Quantum physicists and physics education researchers have the knowledge and experience to devise a QIST educational program that can help prepare students of all ages to participate and lead in the quantum information revolution.

In 2020, NSF and the White House Office of Science and Technology Policy assembled an interagency working group to develop the "Key Concepts for Future Quantum Information Science Learners"—a workshop in which researchers and educators identified core concepts for QIST training, such as qubits, quantum measurement, entanglement, and quantum computing, communication, and sensing. Identifying and building educational resources based on those core concepts will help K–12 and college students become quantum literate and join the growing quantum science and engineering workforce.

In addition to formal education, connecting students with research opportunities, internships, networking, and mentoring is vital for inspiring and nurturing the next generation. To successfully do so, current quantum industry leaders and educators will need to address the lack of representation of women and racial and ethnic minority students and devise approaches to attract demographic groups that are currently underrepresented in the field.

**A common QIST curriculum**

One challenge in teaching QIST is the absence of a common curriculum, in large part because the field is highly interdisciplinary and relatively new. At a [conference](#) in February funded by NSF, educators across academia and industry reflected on the urgent need for bachelor's degree programs, courses, and curricular materials for quantum information science and engineering.

Interdisciplinary bachelor's and master's programs are being developed at universities across the US and Europe. Those programs are based in physics departments, engineering schools, and interdisciplinary centers. Nearly all the programs have at least one course specifically focused on quantum computing or information. The course can rely on traditional textbooks, such as those by [Michael Nielsen and Isaac Chuang](#), [David Mermin](#), or [Benjamin Schumacher and Michael Westmoreland](#), or design the syllabus around resources such as the [open-source Qiskit textbook](#) or other [interactive textbooks](#).

Those courses on quantum computing and information are invariably surrounded by a host of supporting courses designed to meet the highly interdisciplinary demands of QIST, such as programming or computer science foundations, linear algebra, or electrical engineering. Other courses span topics outside core QIST areas or delve into applications related to materials science, chemistry, drug design, machine learning, forecasting, communication, and sensing. Depending on the program's emphasis, it may require different levels of lab experience, internship, and hands-on projects.

**Improving QIST curricula**

Thus far, research on the effectiveness of QIST-related courses, curricula, and pedagogies has been limited in scope (see the article by Chandralekha Singh, Mario Belloni, and Wolfgang Christian, [*Physics Today*, August 2006, page 43](#)). Postsecondary educators have investigated the nature of [students' learning difficulties](#) in quantum courses. Educators have also developed [modular QIST learning tools](#) that can be adapted to courses at the undergraduate and graduate levels.

One such tool is the [Quantum Interactive Learning Tutorials](#) (QuILTs), which are designed to focus on the [topics](#) students have been shown to struggle with, such as the time-evolution of quantum states, quantum measurements, and quantum key distribution. QuILTs use guided inquiry-based learning sequences in which concepts build on one another. Common conceptual difficulties are brought out explicitly by, for example, a written dialogue between two hypothetical students in which one student understands a concept and the other doesn't. Students working through the lesson contemplate which student is correct and why with the help of hints along the way.

Each unit starts with measurable learning objectives and then aligns the instruction and assessment with those objectives. The lessons are often supplemented with computer-based visualization tools. QuILT development is iterative: Data from each [implementation with students](#) are used to improve the tool.

**Academia and industry**

In addition to providing students in-house research experiences in individual faculty research labs or in capstone courses, post-secondary educators should consider partnering with companies and national laboratories to provide students internship and networking opportunities. The Quantum Economic Development Consortium (QED-C), for example, connects students with full-time QIST jobs in companies and national labs. Those opportunities can get students excited about QIST careers and give them a well-rounded understanding of what it is like to work in the area in a non-academic setting.

Partnership between academia and industry can also be immensely helpful for developing degree programs, courses, and curricula that balance academic rigor with the practical needs of the workplace. Since 2016, [IBM Quantum](#) has made its superconducting quantum computers available to the public, and that access enables a hands-on approach to QIST education. Newcomers and experts from around the world can use a free cloud-based interface to run quantum programs on hardware that is located and maintained in a central facility. Other hardware platforms, such as [IonQ](#)'s ion trap–based quantum computer, offer similar cloud-based interfaces.

IBM has also released an open-source software development suite called [Qiskit](#), which is based on the Python programming language, and an open-source [interactive textbook](#) that demonstrates how to use Qiskit to implement quantum algorithms on real quantum systems. By directly designing and running quantum algorithms, students gain a better understanding of the benefits quantum systems will bring to computation, sensing, and communication. The textbook covers both elementary concepts and the state-of-the-art and, with additional resources, can be adapted to a wide range of learners.

IBM has also produced a [series of videos](#) that can readily be integrated into coursework, including a full semester-long course with 27 hours of lectures and labs that introduce quantum information science concepts through the use of [hands-on labs.](#) A typical introductory course on quantum computation can easily integrate those resources by pairing lecture topics with hands-on activities, as shown in Figure 1.

As students learn introductory quantum concepts, they can check their understanding with Qiskit lessons. For example, when they learn about quantum logic gates, they can apply those gates to create a circuit that implements a simple quantum algorithm. Similarly, students learning about quantum algorithms can implement them on simulated or real quantum computers and see how those algorithms solve real problems. Advanced students can opt to learn about the physics behind quantum computer hardware, including how to control the hardware to manipulate quantum states, the physical layout of superconducting circuits, how quantum information is encoded in physical states, and the limitations of existing hardware.

**Hands-on QIST pedagogy**

Hands-on QIST approaches result in deeper understanding and more rapid uptake of quantum concepts, according to a survey conducted after a 2020 remote global summer school hosted by IBM. With more than 4000 students, it was the largest quantum summer school of its kind. The syllabus was based on IBM's educational activities and interactive online textbook.

The attendees were composed of 8% high school students, 40% undergraduates, 30% master's students, and 22% PhD students. We (Asfaw) helped to organize the event and surveyed the 745 attendees before and after the two-week school about their self-reported skill level (see figure 2), on a scale of 0 (no experience with QIST) to 10 (novel research in QIST). Before the summer school (panel **a**), the most frequent answer was no prior experience with QIST, and the distribution was heavily weighted toward a skill level of zero. After the sumer school (panel **b**), the distribution of skill levels shifted to the right by five points.

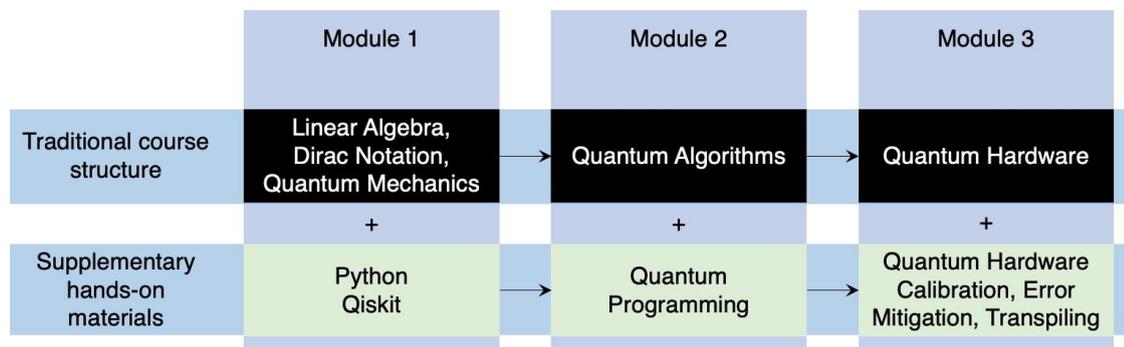

**Figure 1.** IBM recommends supplementing traditional quantum computation course topics with hands-on supplementary materials that incorporate real quantum systems. Adapted from [Qiskit Global Summer School.](#)

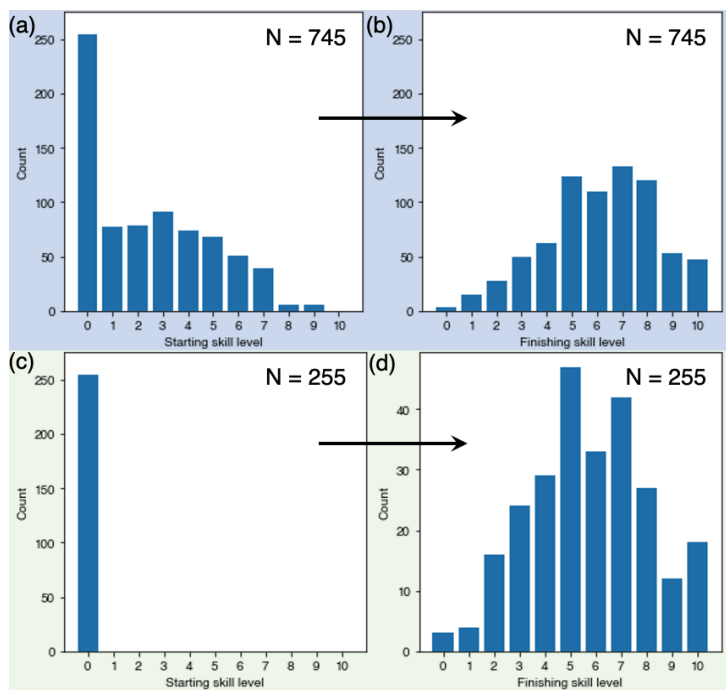

**Figure 2.** IBM's Qiskit Global Summer School in 2020 incorporated traditional and hands-on QIST educational materials. The self-reported skill levels of all the attendees

who responded to the survey before **(a)** and after **(b)** attending the summer school showed about a 5-point increase on average. The distribution for those who entered the summer school with no prior programming experience **(c)** showed the same 5-point shift **(d)**.

Attendees who entered the summer school with no prior programming experience also self-reported on average a five-point increase in their skill level. Although self-reported skill levels do not give a direct measurement of learning outcomes, they do indicate success in lowering the intimidation that students of diverse educational backgrounds feel about learning quantum concepts.

**Upskilling quantum-adjacent roles**

Decades ago, only a select group of specialists had access to and could operate computers. Because of advanced software and layers of abstraction from the underlying semiconductor technology, today everyone can use a computer. At the current stage of the second quantum revolution, working in the industry and development of quantum computers still generally requires a PhD. But a growing set of "quantum-adjacent" industry jobs also require quantum literacy. Those jobs include programming, software development, electronics, cryogenics, vacuum technology, and algorithms.

For those with expertise in quantum-adjacent areas to become familiar with and interested in QIST, courses are needed to help them develop basic understanding of QIST concepts such as quantum bits, superposition, and entanglement. Michael Raymer of the University of Oregon has developed one such quantum literacy course and textbook, *Quantum Physics: What Everyone Needs to Know* (2009), which is appropriate for those interested in quantum-adjacent jobs or undergraduate students with no previous understanding of QIST concepts.

Sophia Economou and Edwin Barnes of Virginia Tech and Terry Rudolph of Imperial College London have developed a [teaching method](#) for college and high school students that uses Rudolph's pictorial representations to minimize the need for linear algebra to explain QIST concepts such as quantum gates that are essential for quantum literacy and quantum-adjacent jobs. In addition, Diana Franklin and Kaitlin Smith from the University of Chicago are offering a sequence of EdX courses that started in January 2021. Requiring only basic algebra skills, their "[Intro to Quantum Computing for Everyone](#)" provides intuitive introductions to quantum physics concepts—such as superposition, measurement, entanglement—that focus on the implications of the quantum operations rather than the details.

**QIST equity**

Introducing students to QIST concepts early is critical to getting them excited about the topic and related careers. That fact is especially true for students from diverse backgrounds, from low-income families, or whose parents did not attend college. Without exposure at the K–12 level, those students will likely not pursue courses related to QIST technologies at the college level.

The [Institute for Quantum Computing at the University of Waterloo, Canada](#), have been conducting professional development activities and disseminating materials to K–12 educators to help them introduce their students to QIST in a conceptual manner. In those lesson plans, for example, students learn about quantum key distribution and the effect of an eavesdropper by analyzing the polarization states of light with inexpensive polarizers.

Postsecondary educators should also consider working with K–12 educators—for example, as part of the new [Quantum for All](#) program—to incorporate QIST concepts in national K–12 standards and in existing courses in physics, chemistry, computer science, or mathematics. They should also think about developing short courses focused exclusively on QIST concepts.

Recently, the American Association of Physics Teachers partnered with IBM to organize K–12 teacher professional development workshops in which teachers learned to use cloud-based quantum computers so that they can incorporate it in their lessons. K–12 educators can also use the book [*Q is for Quantum*](#) (2017) by Terry Rudolph from Imperial College London, which presents diagrammatic explanations of QIST concepts that are accessible to anyone with basic arithmetic skills. The website [Quantum Atlas](#) hosts a collection of images, animations, and audio and written explanation on quantum topics for public and educators alike.

Students from traditionally underrepresented groups need opportunities to access resources if they are to become leaders in QIST. Fortunately, quantum computers are accessible to any school through the cloud. And the recently formed [IBM-HBCU Quantum Center](#) is collaborating with historically Black colleges and universities to develop a more diverse future QIST workforce and advance quantum information science.

Inclusive and equitable approaches to teaching, mentoring, and support are also critical to ensure [equitable outcomes](#) (see the article by Lauren Aguilar, Greg Walton, and Carl Wieman, [*Physics Today,* May 2014, page 43](#)). For example, at the beginning of a quantum course, a short [ecological belonging intervention](#), in which the class discusses how social and academic adversity is normal and surmountable, can help underrepresented students develop a better sense of belonging and thus improve their learning outcomes. Educators, internship or research mentors, course or workshop instructors, and outreach leaders or coordinators should receive [inclusive mentoring training](#) to learn how to increase their mentee's [sense of belonging](#), [self-efficacy](#), and [identity](#) as someone who can excel in QIST.

***Acknowledgement:*** *Singh and Levy acknowledge NSF (PHY-1806691, PHY-1913034) for support.*